\documentclass[12pt]{iopart}
\begin{document}
\title[Proper vertex]{Self-consistent nonperturbative anomalous dimensions}

\author{R Delbourgo\dag}

\address{\dag\ School of Mathematics and Physics, University of Tasmania,
         Box 252-21, AUSTRALIA 7001}
\ead{Bob.Delbourgo@utas.edu.au}

\begin{abstract}
A self-consistent treatment of two and three point functions in models with 
trilinear interactions forces them to have opposite anomalous dimensions.
We indicate how the anomalous dimension can be extracted non-perturbatively
by solving and suitably truncating the topologies of the full Dyson-Schwinger
set of equations. The first step requires a sensible ansatz for the full
vertex part, which conforms to first order perturbation theory at least. We
model this vertex to obtain typical transcendental relations between anomalous 
dimension and coupling constant $g$ which coincide with known results
to order $g^4$.
\end{abstract}

\submitto{\JPA}

\maketitle

\section{Non-perturbative equations}
Renormalizable quantum field theories like QED, QCD or pseudoscalar Yukawa 
theory are characterised by anomalous dimensions which determine the
asymptotic properties of the Green functions as all momenta are scaled.
Specifically the anomalous dimension $\gamma$ of a quantum field is determined
through the asymptotic behaviour of the propagator $(p^2)^{-1-\gamma}$ when
$p^2\rightarrow\infty$. The calculation of $\gamma$ as a power series in
the renormalized coupling constant $g$, arising in the trilinear interaction
Lagrangian, has occupied much time and effort and is known up to order $g^8$ 
for certain models. It would be nice if one could calculate $\gamma$ to all
orders of $g$ by summing subclasses of diagrams, corresponding to particular 
topologies, without excessive effort. This paper is devoted to outlining a
method by which this might be achieved. It is based on conformal scaling of 
Green functions at asymptotic momenta, with anomalous dimensions providing
the self-regulation of the field theory \cite{MT,P} in question. Such a 
proposal makes eminently good sense at a zero $g=g^*$ of the beta function on 
the positive real line and we simply assume that this applies in all that 
follows. However it must be pointed out that for the models considered later 
there is no indication of such a zero and with its corresponding coupling 
constant.

The basic idea behind the method is to eliminate the renormalization
constants as far as it is possible and to write the renormalized Green 
functions in terms of one another as a power series in the coupling constant
with ever more complicated topologies. This produces a set of self-consistent
Schwinger-Dyson equations of the full propagator in terms of itself and the 
proper vertex part, though there exist infinitely many different topological
terms in the skeleton expansion of course. 
In the end one is forced to truncate these topological
contributions, but in any case it is rather easy to show (see below) that the 
only self-consistent behaviour is one where the anomalous vertex and propagator
dimensions are oppositely related. The equations in principle then yield an
equation for the anomalous dimensions in terms of the renormalized coupling 
constant. In the past it has proved relatively simple to sum over particular 
topologies where the vertex function remains undressed, but in this paper we
shall emphasize the importance of considering the full three-point vertex 
(not just in some kinematic limit) when studying these equations; this is the 
novel aspect of our work and by this means one may hope to improve on the 
rainbow \cite{Detal, FKH}, ladder and chain \cite{BK} approximations of 
diagram sets --- the favourite ones studied thus far and with great aplomb by 
Broadhurst and Kreimer.

To illustrate what is involved, consider the case of the renormalizable 
$g\sigma\phi^\dagger\phi$ interaction in six dimensions, rather than
$g\phi^3/6!$ theory \cite{MW}, ignoring 
triple $\sigma$ interactions for the moment. Including renormalization 
constants, the equation for the $\phi$ and $\sigma$ propagators, $\Delta_\phi$ 
and $\Delta_\sigma$ respectively, expressed in terms of full Green functions, 
in the massless case reads
\begin{eqnarray}
 \Delta^{-1}_\phi(p)&=&Z_\phi p^2-ig^2 Z_g\int\bar{d}^6k
     \Gamma(p,p+k,k)\Delta_\phi(k)\Delta_\sigma(p+k)\nonumber\\
&=&Z_\phi p^2-ig^2\!\int\!\bar{d}^6k \Gamma(p,p+k,k)\Delta_\phi(k)
                 \Delta_\sigma(p+k)\Gamma(p,p+k,k)\nonumber \\
&-&\!g^4\!\!\int\!\!\bar{d}^6k\bar{d}^6q\,\Gamma(p,\!p\!+\!k,\!k)\Delta_\phi(k)
 \Delta_\sigma(p\!+\!k)\Delta_\phi(p\!+\!q)\Delta_\sigma(q)
 \Gamma(p,\!p\!+\!q,\!q)\nonumber\\
& &\qquad\qquad\,\Gamma(q,k,q-k)\Delta_\phi(k-q)\Gamma(p+k,q+k,k-q)\cdots
\end{eqnarray}
\begin{eqnarray}
 \Delta^{-1}_\sigma(p)&=&Z_\sigma p^2-ig^2 Z_g\int\bar{d}^6k
     \,\Gamma(p,p+k,k)\Delta_\phi(k)\Delta_\phi(p+k)\nonumber\\
 &=&Z_\sigma p^2-ig^2\!\int\!\bar{d}^6k\,\Gamma(p,p+k,k)\Delta_\phi(k)
                 \Delta_\phi(p+k)\Gamma(p,p+k,k)\nonumber \\
  &-&\!\!g^4\!\!\int\!\!\bar{d}^6k\bar{d}^6q\Gamma(p,p\!+\!k,k)\Delta_\phi(k)
  \Delta_\phi(p\!+\!k)\Delta_\phi(p\!+\!q)\Delta_\phi(q)
  \Gamma(p,p\!+\!q,q)\nonumber\\
 & &\qquad\qquad\Gamma(q,k,q-k)\Delta_\sigma(k-q)\Gamma(p+k,q+k,k-q)\cdots
\end{eqnarray}
For each field, we know that in renormalized perturbation theory the sum of the 
series will produce the large $p$ scaling behaviour, $\Delta^{-1}(p)\simeq 
cp^2(-p^2/\mu^2)^\gamma$, where $\gamma$ is given as a series in $g^2$, which 
can be computed \footnote{We anticipate that the constant $c$ appearing in 
$\Delta$ behaves like unity when $\gamma\rightarrow 0$; this is 
because the imaginary part of $\Gamma(1+\gamma)/(p^2+i\epsilon)^{1+\gamma}$ 
equals $-\pi\theta(-p^2)/\Gamma(-\gamma)(p^2)^{1+\gamma}$ and this generalised
function reduces to $-\pi\delta(p^2)$ for the free case $\gamma=0$.}
order by order in perturbation theory, albeit with greater and greater 
pain as the power of $g^2$ rises. The purpose of this paper is to 
look for relations between $\gamma$ and $g^2$, which correspond to particular 
truncations of various topological contributions to the self-energy; such 
relations are normally given by a transcendental equation that can be found via
the Dyson-Schwinger equations. But how do we find them by suitably manipulating
eqs (1) and (2)? For the moment neglect the first term on the right, which 
involves the renormalization constant and seems to scale as $p^2$; the 
remaining terms have the form
$$g^2F_2(p,\mu,\gamma(g^2)) + g^4F_4(p,\mu,\gamma(g^2)) + \cdots$$
and we must ask how they sum up to the scaling form on the left-hand side.
It is possible, but extremely unlikely, that each of these terms behaves as 
$p^2$ aside from logarithms and that each contains an infinity (which is 
subtracted off by the $Z$-factor on the right) and that they somehow combine to 
produce the anomalous scaling; if that were true there would be little point
in using the skeleton expansion as a means of transcending perturbation theory.
A more likely scenario, which we shall {\em assume} hereafter, is that each of 
the contributions $F_N$ scale in exactly the same way. Then it is not hard to 
work out what must be the scaling behaviour of the proper vertex $\Gamma$ to 
produce $(p^2)^{1+\gamma}$ at large $p^2$. One easily establishes that
\begin{equation}
  \Gamma(\lambda p,\lambda(p+k),\lambda k)\sim\lambda^{2\gamma_\Gamma}
  \Gamma(p,p+k,k); \qquad \gamma_\Gamma = \gamma_\phi+\gamma_\sigma/2
\end{equation}
will ensure that {\em all} of the topological contributions $F_N$ to the 
two-point function yield the same scaling as the external momentum 
$p\rightarrow\infty$. [Without losing too much generality one may fix the 
vertex function $\Gamma=1$ either at the symmetrical Euclidean
point $p^2=(p+k)^2=k^2=-\mu^2$ {\bf or} at zero-momentum transfer $k=0, p^2
= (p+k)^2=-\mu^2$, but that still leaves vast freedom in the dynamics of 
$\Gamma$ in (2), through its dependence on the momentum ratios.] 
As for justifying why we drop the renormalization terms $Z p^2$, we 
note that the wave function renormalization constant may be generally defined 
by $Z^{-1}\equiv \lim_{p^2\rightarrow\infty} p^2\Delta(p)$ and then $Z$ can 
well vanish \footnote{The condition $Z=0$ applies to composite 
particles and agrees with this observation, but becomes problematic for 
positive $\gamma$.} for negative $\gamma$, in which case dropping it from 
the self-consistent but nonperturbative Schwinger-Dyson equations is not 
entirely absurd in the asymptotic limit. 

Specifically, let us contemplate the situation where one the fields, say 
$\sigma$, remains undressed (some sort of quenched approximation in which 
closed $\phi$-loop graphs are disregarded, signifying that induced multisigma
interactions can be dropped). Thus its corresponding $Z_\sigma=1, \gamma_\sigma
=0$, so the scaling of $\Gamma$ just becomes tied to the asymptotic behaviour 
of the field $\phi$: they are inverse to one another. This connection is 
unsurprising especially in QED where the Ward identity 
$\Gamma(p,p,0)=\partial\Delta^{-1}/\partial p$ at zero photon momentum, 
already indicates that the scaling properties of the charged field and the 
soft vertex are intimately tied (`$Z_1=Z_2$' in a perturbative context). 
In Section 2 we summarize what is known about the vertex function in $\sigma
\phi^\dagger\phi$ and Yukawa theory, the two models upon which we focus. There 
we look for a nonperturbative version of $\Gamma$ which reduces to first order 
perturbation theory, {\em with the correct singularities}. Then in Section 3 we 
show how this can be used to determine the relation between $\gamma$ and $g^2$ 
even in the simplest kind of truncation encompassing the chain and rainbow
summations with various model vertex functions.

\section{Structure of the proper vertex} 

\subsection{Lowest order structure}

One of the most elementary exercises in quantum field theory is to work out the 
first order corrections to the propagator and vertex part, where we normally 
encounter infinities that must be renormalized. Working in 2$\ell$ dimensions
(for isolating the infinities) leads one to the typical expressions for the 
self-energy $\Sigma$ and proper vertex correction $\Lambda$,
$$\Sigma(p)=c_2g^2\int\!\int_0^1\!dx\,dy\,
   \frac{\delta(x+y-1)\,\Gamma(2-\ell)}{[p^2 xy-m_1^2x-m_2^2y]^{2-\ell}}$$
$$\Lambda(p_1,p_2,p_3)=c_3g^2\int\!\int\!\int_0^1\!dx\,dy\,dz\,
  \frac{\delta(x+y+z-1)\,\Gamma(3-\ell)}
  {[p_1^2yz+p_2^2zx+p_3^2xy-m_1^2x-m_2^2y-m_3^2z]^{3-\ell}},$$
where $\ell\rightarrow n=2$ or 3 in the integer limit and $c_i$ are symmetry 
factors including $1/(4\pi)^\ell$.  Subtraction of the pole terms 
by renormalization constants produces logarithms of the denominators of each of 
the right hand sides, normalized at some mass scale $\mu$, which is how 
dimensional transmutation and the renormalization prescription enters. One can 
recognize the resulting expressions as the first terms of an expansion ---
which strictly makes sense at a zero of the beta-function when we proceed to 
the nonperturbative regime --- of the Green functions in powers of $g^2$; 
for example for the massless two-point 
function, where the logarithmic term is easily computed, we end up 
straightforwardly with 
$$\Delta^{-1}(p)=p^2+c_2g^2p^2\log(-p^2/\mu^2)+\cdots \simeq 
  p^2(-p^2/\mu^2)^\gamma$$
in keeping with $\gamma = c_2g^2 + \cdots$. For the three-point function,
renormalized at zero $\sigma$ momentum, the logarithm is actually a 
dilogarithm, being given by the integral
\begin{eqnarray}
-\Lambda&=&\int_0^1\!\int_0^1\!\int_0^1\!dx\,dy\,dz\, \delta(x+y+z-1)
 \log\left(\frac{p_1^2yz+p_2^2zx+p_3^2xy}{-\mu^2(yz+zx)}\right)\nonumber\\
 &=&\!\frac{c_3g^2}{2}\!\!\int_0^1\!\!\sigma d\sigma\!\!\int_{-1}^1\!\!\!d\tau
 \log\left[ \left(\frac{-p_1^2}{2\mu^2}\right)(1\!-\!\tau)\!+\!
 \left(\frac{-p_2^2}{2\mu^2}\right)(1\!+\!\tau)\!-\!
 \frac{p_3^2\sigma(1\!-\!\tau^2)}{4\mu^2(1\!-\!\sigma)}\right]
\end{eqnarray}
whose singularities characterize the `triangle' graph. It simplifies to pure
logarithms when $p_3^2\rightarrow 0$, being proportional to
$$\frac{1}{2}\left[\log\left(\frac{p_1^2p_2^2}{\mu^4}\right) - 2 
+\frac{p_2^2+p_1^2}{p_2^2-p_1^2}\log\left(\frac{p_2^2}{p_1^2}\right)\right],$$
which one should note is the order $\gamma$ term of the expression
$$\frac{\mu^2}{(1+\gamma)(p_1^2-p_2^2)}\left[\left(-\frac{p_2^2}
{\mu^2}\right)^{\gamma+1}-\left(-\frac{p_1^2}{\mu^2}\right)^{\gamma+1}\right].$$
This bears an uncanny resemblance to the ratio $[\Delta^{-1}(p_2)-
\Delta^{-1}(p_1)]/(p_2^2-p_1^2)$ that one encounters when 
``solving'' for the longitudinal 3-point vertex in gauge theories; we'll come 
back to this point presently. In any case it strongly suggests that the 
nonperturbative form of the vertex possessing triangular topology exponentiates 
to
\begin{equation}
\Gamma(p_1,p_2,p_3)=2\int\!\!\int\!\!\int_0^1\!dx\,dy\,dz\,\rho\delta(x+y+z-1)
 \left[\frac{p_1^2yz+p_2^2zx+p_3^2xy}{-\mu^2(yz+zx)}\right]^{\gamma_\Gamma},
\end{equation}
where the (symmetric in $y,z$) spectral function $\rho(x,y,z)$ equals 1 up to 
first order in $g^2$ and $\gamma_\Gamma=-c_3g^2/2$. If we restrict ourselves to 
such topology, neglect the dressing of $\sigma$ and reinsert (4) into (1), this 
will produce a self-consistent equation for $\gamma_\phi$ in terms of $g^2$, 
as we already know that uniform scaling of $F_N$ requires $\gamma_\Gamma
=\gamma_\phi$ when $\sigma$ is quenched. 

The Feynman parametric form of $\Lambda$ in the limit as $p_3^2 = 0$, (but 
{\em not}\, $p_3=0$) is very suggestive of another Lehmann-like representation
which one encounters in the gauge technique. Thus by changing variable to
$w^2 = m_1^2/y + m_2^2/x$ in the self-energy, one may convert from Feynman to 
dispersive form:
$$\Sigma(p) = \int_{(m_1+m_2)^2}^\infty\, dw^2\,\sigma(w^2)/(p^2-w^2).$$
In the particular case that $m_1=m_2=m$ for the vertex function, which is often 
true, one can make a similar conversion in $\Lambda$ as $p_3^2\rightarrow 0$:
$$\Lambda\propto\int_0^1 dx\int_{-x}^xdu\frac{\frac{1}{2}\Gamma(3-\ell)}
 {[\frac{1}{2}(p_1^2+p_2^2)x(1-x)+\frac{1}{2}
 (p_1^2-p_2^2)u(1-x)-m^2x]^{3-\ell}}$$
$$=\int_0^1 \frac{\Gamma(2-\ell)\,dx}{(p_2^2-p_1^2)(1-x)}\left[
 \frac{1}{(p_1^2x(1-x)-m^2x)^{2-\ell}}-\frac{1}{(p_2^2x(1-x)-m^2x)^{2-\ell}}
 \right],$$
$${\rm and~closely~resembling}\quad\frac{\Sigma(p_1^2)- \Sigma(p_2^2)}
{p_2^2-p_1^2}=\int\frac{dw^2\,\sigma(w^2)}{(p_1^2-w^2)(p_2^2-w^2)}.$$
Thus we anticipate that the vertex possesses a simplified dispersive-like
representation when one of the momenta is lightlike, and this property might 
prove rather useful. Further progress may need additional truncation of the 
skeleton expansion and other practical simplifications.

\subsection{Perturbation theory}

These remarks apply to all theories with basic trilinear interactions. Before
looking at nonperturbative aspects of the models $g(\sigma\phi^\dag\phi)_{6D}$
and $(g\bar{\psi}\gamma_5\psi\phi)_{4D}$, let us first note the order $g^2$ 
results for $c_2,c_3$ since they supply helpful weak-coupling limits
for the two and three point functions involving massless fields $\phi, 
\sigma$ and $\psi$. 

In $g(\phi^\dag\phi\sigma)_{6D}$ one trivially finds that
\begin{equation}
 \Sigma(p)=\frac{g^2}{p^4}\left(\frac{-p^2}{4\pi}\right)^\ell
  \frac{\Gamma(2-\ell)\Gamma(\ell-1)\Gamma(\ell-1)}{\Gamma(2\ell-2)}
\end{equation}
where the limit $\ell\rightarrow 3$ must be taken. Renormalizing at
$p^2=-\mu^2$, one ends up with
$$\Sigma(p) = -\frac{g^2p^2}{6(4\pi)^3}\ln\left(-\frac{p^2}{\mu^2}\right).$$
Likewise the vertex correction is just expression (4), with $c_3=1/(4\pi)^3$.
It may be expressed \cite{DB} as an Appell function, but better still, it can 
be converted into a symmetrical sum of three hypergeometric functions in
arbitrary dimensions \cite{D2}; we will utilise this presently.

Turning to $(g\bar{\psi}\gamma_5\psi\phi)_{4D}$ theory,
the inverse renormalized $\psi$ propagator to that order reads,
\begin{equation}
 S^{-1}(p) = \gamma\cdot p\left[1 - \frac{g^2}{32\pi^2}\log\left(-
             \frac{p^2}{\mu^2}\right) +\ldots\right],
\end{equation}
while the fully off-shell proper vertex part correction is given by
\begin{equation}
\Lambda_5(p',p) = (A + B[\gamma\cdot p',\gamma\cdot p])\gamma_5,
\end{equation}
where, of the two scalar functions $A$ and $B$, only the former carries the 
ultraviolet divergence. (A massive theory, apart from modifying $A$ and
$B$, would have led to further terms like 
$[C\gamma\cdot p + C'\gamma\cdot p']\gamma_5$.)  One readily finds that
\begin{equation}
A=\frac{ig^2}{2}\int \frac{\bar{d}^{2\ell}k}{k^2}\left[\frac{1}{(p+k)^2}
  +\frac{1}{(p'+k)^2}-\frac{(p-p')^2}{(p+k)^2(p'+k)^2} \right]
\end{equation}
and
\begin{eqnarray}
B&=&-\frac{ig^2}{2\Delta}\int\frac{\bar{d}^{2\ell}k}{k^2}\left[
 (p^2-p'^2)\left(\frac{1}{(p+k)^2}-\frac{1}{(p'+k)^2}\right)+
 (p-p')^2\times\right. \nonumber \\
& &\left. \left(\frac{p^2+p'^2-(p-p')^2}{(p\!+\!k)^2(p'\!+\!k)^2}\!+\!
 \frac{1}{(p\!+\!k)^2}+\frac{1}{(p'\!+\!k)^2}-
 \frac{2k^2}{(p\!+\!k)^2(p'\!+\!k)^2} \right)\right]
\end{eqnarray}
to be taken in the limit $\ell\rightarrow 2$. In eq. (10), $\Delta$ is
nothing but the K\"allen function, namely
$$\Delta\equiv 4[(p.p')^2-p^2p'^2]=p^4+p'^4+(p-p')^4-2p^2p'^2-2p^2(p-p')^2
 -2p'^2(p-p')^2.$$
It is interesting to look at the case $(p-p')^2\rightarrow 0$ in (9) and (10) 
before worrying about renormalization; one gets
$$\Lambda_5\rightarrow \frac{ig^2}{2}\int \frac{\bar{d}^{2\ell}k}{k^2}
 \left[\left(\frac{1}{(p'+k)^2}+\frac{1}{(p+k)^2}\right) -
 \frac{[\gamma\cdot p',\gamma\cdot p]}{p'^2-p^2}
 \left(\frac{1}{(p'+k)^2}-\frac{1}{(p+k)^2}\right)\right]\gamma_5.$$
Again we notice the structure $(\Sigma(p')-\Sigma(p))/(p'^2-p^2)$ in the
finite part, even though we are not dealing with a gauge theory. Anyhow,
renormalizing so that $\Lambda_5(p,p)=0$ for $p^2=-\mu^2$, the finite $B$-type 
term remains unaffected, and we are left with the lightlike limit
$$\Gamma_5\rightarrow \gamma_5 - \frac{g^2}{32\pi^2}\left[ 
  \log\left(\frac{p^2p'^2}{\mu^4}\right) + \frac{\ln(p'^2/p^2)}{p^2-p'^2}
  [\gamma\cdot p',\gamma\cdot p] \right]\gamma_5.$$
Our aim is to identify these expressions as first order in $g^2$ parts of
some nonperturbative construct, so as to capture at the very least the
triangular topology of the full vertex function, and then see how far we
can take it from there.

\subsection{Nonperturbative form}

Let us start by reducing the perturbative vertex to a more manageable form.
The first order expression for $\Lambda(p_1,p_2,p_3)$ involves hypergeometric
functions of two variables when the dimension and masses are arbitrary. 
However it can be simplified for $m_i=0$ in a very elegant way and Davydychev 
\cite{D2} has shown the way to do this: draw three lengths of size $\sqrt{p_1^2},
\sqrt{p_2^2},\sqrt{p_3^2}$ and let $\Theta_{123}=1$ if a Euclidean triangle 
can be drawn with those sides, 0 otherwise. Then $\Delta_E= -\Delta =
2(p_1^2p_2^2+p_2^2p_3^2+p_3^2p_1^2)-p_1^4-p_2^4-p_3^4$ is four times the square
of the area of such a triangle. The internal angles $\phi_i$ of the triangle 
are given in an obvious notation by $2\sin\phi_1=\sqrt{\Delta_E/p_2^2p_3^2},\,
2\cos\phi_1 = (p_2^2+p_3^2-p_1^2)/\sqrt{p_2^2p_3^2}$, etc. and
of course $\phi_1+\phi_2+\phi_3=\pi$. The result for
$$\Lambda(p_1^2,p_2^2,p_3^2)=ig^2\int\bar{d}^{2\ell}r
  /[r^2(p_1-r)^2(p_2+r)^2],$$
which leads to the hypergeometric answer,
\begin{eqnarray}
 \Lambda&=&\frac{g^2\Gamma(2-\ell)}
{(4\pi)^\ell(-p_1^2p_2^2p_3^2)^{2-\ell}}\left[
 2\pi\Delta_E^{3/2-\ell}\Theta_{123}-\frac{\Gamma^2(\ell-1)}{\Gamma(2\ell-2)}
 \times\right.\nonumber \\
& &\left.\qquad\left(\frac{(p_1^2p_2^2)^{2-\ell}}{p_1^2+p_2^2-p_3^2} 
 {_2F_1}(1,\frac{1}{2};\ell\!-\!\frac{1}{2};-\frac{\Delta_E}
 {(p_1^2\!+\!p_2^2\!-\!p_3^2)^2})+\!{\rm 2~perms}\right)\right]
\end{eqnarray}
can be converted into the beautiful form,
\begin{equation}
\Lambda=\frac{g^2\Gamma(2\!-\!\ell)\Delta_E^{3/2-\ell}}
         {(4\pi)^\ell(-p_1^2p_2^2p_3^2)^{2-\ell}}\left[2\pi\Theta_{123}-
 \frac{\Gamma^2(\ell\!-\!1)}{\Gamma(2\ell\!-\!3)}\sum_{k=1}^3\!
 \int_0^{2\phi_k}\!\!\!d\chi\,(4\sin^2\frac{\chi}{2})^{\ell-2}\right].
\end{equation}
Another useful expression (in that limits can be taken more easily) is obtained 
by adopting an integral representation of the hypergeometric functions; thus
\begin{equation}
 \Lambda=\frac{g^2\Gamma(2\!-\!\ell)\Delta_E^{3/2-\ell}}
         {(4\pi)^\ell(-p_1^2p_2^2p_3^2)^{2-\ell}}\left[2\pi\Theta_{123}\!-\!
 \frac{\Gamma^2(\ell\!-\!1)}{\Gamma(2\ell\!-\!3)}\!\sum\!\!
 \int_0^1\!\!\!\frac{(p_i^2p_j^2)^{2-\ell}t^{\ell-5/2}\,dt}
 {\Delta_E^{3/2-\ell}\!\sqrt{4p_i^2p_j^2\!-\!\Delta_Et}}\right]\!.
\end{equation}
Davydychev has gone further and expressed $\Lambda$ as a series in 
$(\ell\!-\!2)$ leading to polylog functions Ls$_j$, but we shall not require 
that expansion. Note that the residue at $\ell=2$ of eq.(12) or (13) vanishes 
(as it must since the vertex is convergent in 4-D), and that the residue in 
6-D at $\ell=3$ reduces to $-g^2/128\pi^3$, because of the identity,
$$p_3^2(p_1^2+p_2^2-p_3^2)+p_1^2(p_2^2+p_3^2-p_1^2)+p_2^2(p_3^2+p_1^2-p_2^2)
 = \Delta_E.$$
Another interesting situation arises when $p_1^2\equiv p^2, 
p_2^2\equiv p'^2, p_3^2=0$, whereupon $\Delta_E=-(p^2-p'^2)^2, \tan^2\phi_1=-1,
\tan^2\phi_2=-1, \tan^2\phi_3=(p^2-p')^2/(p^2+p'^2)^2, \Theta_{123}=0$. Thus
\begin{equation}
 \Lambda(p^2,p'^2,0)=\frac{16g^2\Gamma(2-\ell)\Gamma(1/2)}
 {(16\pi)^\ell\Gamma(\ell-3/2)}
 \frac{[(-p'^2)^{\ell-2}-(-p^2)^{\ell-2}]}{[p'^2-p^2]},
\end{equation}
which allows one to take the limit $p'^2=p^2\rightarrow -\mu^2$, namely
$$\Lambda(-\mu^2,-\mu^2,0)=\frac{16g^2\Gamma(3-\ell)\Gamma(\ell-2)\Gamma(1/2)
   \mu^{2\ell-6}}{(16\pi)^\ell\Gamma(\ell-3/2)}\rightarrow
   -\frac{g^2}{2(4\pi)^3(\ell-3)},$$
as $\ell\rightarrow 3$.

We now suggest a way of ``going nonperturbative", which captures the essence
of the vertex triangular topology. We firstly observe that the asymptotic
form of the inverse propagator $\Delta_\phi^{-1}(p)\simeq (p^2)^{1+\gamma_\phi}$
can be gotten directly from the self-energy $\Sigma(p)$ in $2\ell$ dimensions
simply be making the replacement $\ell = 3+\gamma_\phi$ in the dimensionally
continued result, apart from an overall factor that must be carefully
chosen to accord with the renormalization condition. This procedure will lead 
from (6) to $\Delta_\phi^{-1}(p)=p^2(-p^2/\mu^2)^{\gamma_\phi},$ with 
$\gamma_\phi=-g^2/6(4\pi)^3$ to first order.

Applying the same procedure to the $(\phi^\dag\phi\sigma)_{6D}$ vertex, we will 
end up with
\begin{equation}
\hspace{-1in}\Gamma(p_1,\!p_2,\!p_3)=\frac{p_1^2p_2^2p_3^2}{\Delta_E^{3/2}}
       \left(\frac{p_1^2p_2^2p_3^2}{\mu^2\Delta_E}\right)^{\gamma_\Gamma}
 \!\!\left[2\pi\Theta_{123}\!\!-
 \frac{\Gamma^2(2\!+\!\gamma_\Gamma)}{\Gamma(3\!+\!2\gamma_\Gamma)}
 \sum_{k=1}^3\!
 \int_0^{2\phi_k}\!\!\!d\chi\,(4\sin^2\frac{\chi}{2})^{1+\gamma_\Gamma}\right],
\end{equation}
where $\gamma_\Gamma=-g^2/2(4\pi)^3$ to first order. Realising that 
$\sum_k\phi_k=\pi$ and
$$\sum_{k=1}^3 \sin(2\phi_k) = \Delta_E^{3/2}/2p_1^2p_2^2p_3^2$$
we can readily establish that $\Gamma\rightarrow 1$ when $\gamma_\Gamma
\rightarrow 0$. Also one may check that an expansion of (15) to order $g^2$
reproduces perturbation theory, including terms like 
$$\sum_{k=1}^3 \int_0^{2\phi_k}\!d\chi(4\sin^2\frac{\chi}{2})\log
(4\sin^2\frac{\chi}{2}),$$
which appear after renormalization. Thus we are emboldened to regard (15)
as a decent nonperturbative vertex that incorporates triangular topological
contributions, but whether we can easily make use of it is entirely another 
matter since its form is analytically complicated. Perhaps a more amenable
form of (15) is
\begin{equation}
\hspace{-1in}
\Gamma(p_1^2,p_2^2,p_3^2)=\!\frac{(p_1^2p_2^2p_3^2)^{1+\gamma_\Gamma}}
 {\mu^{2\gamma_\Gamma}\Delta_E^{3/2+\gamma_\Gamma}}
 \frac{\Gamma(3\!+\!2\gamma_\Gamma)}{\Gamma^2(2\!+\!\gamma_\Gamma)}
 \left[2\pi\Theta_{123}\!-\!
 \sum_k\!\!\frac{(\sin\phi_k)^{3\!+\!
2\gamma_\Gamma}}{\cos\phi_k}\!\int_0^1\!\!
 \frac{(1-t)^{1+\gamma_\Gamma}\,dt}{\sqrt{t}(1\!+\!t\tan^2\phi_k)} 
 \right],
\end{equation}
because, without too much trouble, it allows us to take the lightlike limit,
\begin{equation}
 \Gamma(p^2,p'^2,0)= \frac{\mu^2}{(1+\gamma_\Gamma)(p'^2-p^2)}\left[
 \left(-\frac{p^2}{\mu^2}\right)^{1+\gamma_\Gamma}-
 \left(-\frac{p'^2}{\mu^2}\right)^{1+\gamma_\Gamma}\right]\!,
\end{equation}
a result which we foresaw earlier. The special limit when one leg carries
zero momentum, $\Gamma(p^2\!,p^2\!,0)\!=\!(-p^2/\mu^2)^{\gamma_\Gamma}$
is then readily found. Also the renormalized perturbative answer stated at the 
end of the last subsection falls out upon expansion to first order in 
$\gamma_\Gamma$, provided the anomalous dimension is correctly identified.

Let us perform a similar procedure on $(\bar{\psi}\gamma_5\phi\psi)_{4D}$. Here
the full vertex, consists of two terms:
$$\Gamma_5(p',p)=(\Gamma_A+\Gamma_B [\gamma\cdot p',\gamma\cdot p])\gamma_5,$$
whose first order in $g^2$ terms are summarised in (9) and (10). In making the
substitution $\ell\rightarrow 2+\gamma$ so as to obtain a nonperturbative
expression, it is very easy to handle the propagator (7) and arrive at
\begin{equation}
 S^{-1}(p)= \gamma\cdot p (-p^2/\mu^2)^{\gamma_\psi};\qquad
            \gamma_\psi=-g^2/32\pi^2.
\end{equation}
Also it is possible to exponentiate the (renormalized) self-energy like terms in
the vertex parts $A$ and $B$ arising in eqs (9) and (10):
$$1+ig^2\int\bar{d}^{2\ell}k/k^2(p+k)^2 \rightarrow 1+(g^2/16\pi^2)
 \log(-p^2/\mu^2) \rightarrow (-p^2/\mu^2)^{g^2/16\pi^2}.$$
On the other hand the full triangular topology integral (12) or (13)
will produces a {\em finite} result
$$\Gamma_F(p_1^2,p_2^2,p_3^2) = \frac{4}{\sqrt{\Delta_E}}
 \left( \frac{-p_1^2p_2^2p_3^2}{\mu^2\Delta_E}\right)^{\gamma_\Gamma}
 \left[2\pi\Theta_{123}-\frac{\Gamma^2(1+\gamma_{\Gamma_E})}
   {\Gamma(1+2\gamma_{\Gamma_E})}\sum_k\int_0^{2\phi_k}d\theta(
     \sin^2\theta)^{-\gamma_\Gamma} \right] .$$
One may verify that the vertex scaling
behaviour is reproduced in $\Gamma_F$ and that it vanishes for $\gamma_\Gamma
= 0$, as it should in zeroth order.  Anyhow, combining the terms, we arrive
at the nonperturbative Yukawa vertex parts,
\begin{equation}
2\Gamma_A(p',p) = (-p^2/\mu^2)^{\gamma_{\Gamma}}+
                  (-p'^2/\mu^2)^{\gamma_{\Gamma}}-
                  (p-p')^2\Gamma_F
\end{equation}
\begin{equation}
\Delta\Gamma_B(p',p) = (p'^2-p^2)[(-p'^2/\mu^2)^{\gamma_{\Gamma}}-
                  (-p^2/\mu^2)^{\gamma_{\Gamma}}]
                  +(p-p')^2\Gamma_{FB},
\end{equation}
where $\Gamma_{FB}=[(p-p')^2-p^2-p'^2]\Gamma_F
 + 2(-(p-p')^2/\mu^2)^{\gamma_\Gamma}-(-p^2/\mu^2)^{\gamma_\Gamma}
 - (-p'^2/\mu^2)^{\gamma_\Gamma}$.

\section{Applications}
The question is how to apply all this. One's first inclination is to substitute
the vertex (15) or (16) into (1) and (2) so as to find the relation between 
anomalous dimension and coupling constant --- a relation which is normally 
found by a tedious process of perturbative renormalization and is worked out to 
order $g^6$ at least. Our proposal is that nonperturbative forms of propagator 
and vertex avoid the need for renormalization since the skeleton expansion is 
automatically regularized at the physical dimension $D=4$ or 6; as we shall see,
such a procedure will lead to a transcendental relation between $\gamma$ and 
$g^2$. This happens even when one truncates to the first contribution to the 
self-energy, having the form 
$$\int\Gamma(p,p+k)\Delta(p+k)\Delta(k)\Gamma(p+k,p)\,d^Dk,$$
although one ought properly to consider the entire series of skeleton terms 
with their ever more intricated topologies (and matching vertices). Nonetheless,
considering even the first term of the skeleton series is a substantial 
improvement on past efforts \cite{Detal, FKH} and is worthy of study.

The task of evaluating the first term $g^2F_2(p,\mu,\gamma(g^2))$ is rather
daunting: one is required to integrate the product of two dressed 
propagators with the {\em square} of expression (15) or expression (16), and
this is technically very demanding. Numerical calculations are useless from that
viewpoint because one is interested in obtaining the analytical connection 
between $\gamma$ and $g^2$, even if comparison with perturbation theory 
eventually necessitates a power series expansion in $g^2$. A number of helpful 
auxiliary integrals are collected in the appendix, but we are forced to admit 
that the full-blown integration producing $F_2$ is presently beyond our 
technical reach. {\em Faute de mieux} we are forced to approximate the skeleton 
Feynman integral by one which is doable and which captures the essence of the 
idea: the main thing is to ensure that the non-perturbative vertex has the 
correct scaling behaviour, symmetry properties and analytic behaviour as far as 
possible, and that it should by itself regularize the intermediate momentum 
integral.

In the following we shall attempt to use a few approximations to the 
nonperturbative triangular vertex that reflect its main features, in order to 
extract the relation between anomalous dimension and coupling constant. 
Thus we will examine a number of models in which $F_2$ is free of infinities 
and is automatically regularized --- it all too easy to construct models which 
have the correct scaling property but which nevertheless contain infinities ---
but where the vertex singularities are not quite correct.

\subsection{Model 1 for $(\phi^\dag\phi\sigma)_{6D}$}
Here we make the choice $\Gamma(p,p+k,k)=(p^2(p+k)^2/\mu^2k^2)^\gamma$ and
leave the $\sigma$ propagator undressed. This has the virtue of simplicity;
it possesses symmetry at the $\phi$ legs and correct scaling but is otherwise 
awry in its analytical properties and especially its vertex singularities.
Ignoring these defects, and using the results in the Appendix, we obtain the
self-consistency relation
$$ p^{2(1+\gamma)} = ig^2\int\frac{\bar{d}^6k}{k^2(p+k)^{2(1+\gamma)}}
   \left(\frac{p^2(p+k)^2}{k^2}\right)^{2\gamma}, $$
or
\begin{equation}
 1 = a\frac{\Gamma(-1+\gamma)\Gamma(2-2\gamma)\Gamma(2+\gamma)}
     {\Gamma(4-\gamma)\Gamma(1+2\gamma)\Gamma(1-\gamma)};\qquad 
     a\equiv \frac{g^2}{(4\pi)^3}.
\end{equation}
This may be contrasted with the rainbow approximation where the 
self-consistency relation instead reads,
$$p^4\Delta(p) = -ig^2\int \Delta(p+k)\,\bar{d}^6k/k^2 \quad {\rm and}
 \quad 1 = a/\gamma(\gamma-1)(\gamma-2)(\gamma-3).$$
To obtain a perturbative expansion of (21) we take a series in $\gamma$ or $a$ 
as needed to arrive at
$$\gamma_{\rm model 1} = -\frac{a}{6} + \frac{11a^2}{6^3} - \frac{134a^3}{6^5} 
+ \cdots , $$
compared with \cite{BK}
$$\gamma_{\rm rainbow}=-\frac{a}{6}+\frac{11a^2}{6^3}-\frac{206a^3}{6^3}
  +\cdots,\quad
  \gamma_{\rm chain} = -\frac{a}{6}+\frac{11a^2}{6^3}-\frac{170a^3}{6^3}
  +\cdots.$$

\subsection{Model 2 for $(\phi^\dag\phi\sigma)_{6D}$}
We now consider a vertex which better captures the analytical behaviour of
the triangular topology but which is necessarily more complicated than the
previous model. Here we try to mimic some of the dependence on
$\tan^2\phi_i$ which arises in (11) by letting
$$\Gamma^2(p_1^2,p_2^2,p_3^2)=[(p_1^2)^{1+2\gamma}(p_2^2+p_3^2-p_1^2) +
 (p_2^2)^{1+2\gamma}(p_3^2+p_1^2-p_2^2)+p_3^2(p_1^2p_2^2)^\gamma
 (p_1^2+p_2^2-p_3^2)]/\Delta_E.$$
Making use of the equations in the appendix, we may arrive at the 
self-consistency relation,
\begin{eqnarray}
 1&=&\frac{a}{6}\left[\frac{1}{(2-\gamma)(\gamma-1)}-\frac{1}{\gamma(1-\gamma)}
    + \frac{1}{(2+\gamma)(3+\gamma)}-\frac{1}{(1+\gamma)(2+\gamma)}\right]
  \nonumber \\
&=& -\frac{a}{6}\left[\frac{1}{\gamma}+\frac{11}{6}+\frac{41\gamma}{36}
 +\cdots \right].
\end{eqnarray}
This corresponds to the series
$$\gamma_{\rm model 2} = -\frac{a}{6} + \frac{11a^2}{6^3} - \frac{162a^3}{6^5} 
+ \cdots $$

\subsection{Model 3 for $(\bar{\psi}\gamma_5\psi\phi)_{4D}$}
Before making any approximations, we may note that the $F_2$ contribution
to the inverse fermion propagator $\gamma\cdot p S^{-1}(p)$, with 
$$\Gamma_5=(\Gamma_A + \Gamma_B[\gamma\cdot p,\gamma\cdot p'])\gamma_5$$
can be written as
\begin{equation}
\hspace{-0.5in}
 i\frac{g^2}{p^2}\int\frac{\bar{d}^4k}{((p+k)^2)^{1+\gamma}}\Delta_\sigma(k)
 \left[\frac{1}{2}(p^2+(p+k)^2-k^2)(\Gamma_A^2-\Delta_E\Gamma_B^2)
       +\Delta_E\Gamma_A\Gamma_B\right].
\end{equation}
Therefore, if we use a quenched $\sigma$ approximation, the $F_2$ term leads
to the following self-consistent relation for the anomalous dimension,
\begin{equation}
\hspace{-0.8in}
 (p^2)^{1+\gamma}= ig^2\int\frac{\bar{d}^4k}{2k^2((p+k)^2)^{1+\gamma}}
 \left[(p^2+(p+k)^2-k^2)(\Gamma_A^2-\Delta_E\Gamma_B^2)
       +2\Delta_E\Gamma_A\Gamma_B\right]
\end{equation}
into which we may feed various models for the vertex parts $\Gamma_{A,B}$.
Of course we would dearly have loved to make use of their nonperturbative
forms given in the previous section (consistent with triangular topology) but 
find the resulting computation too hard analytically; so we are obliged to 
model something resembling the true vertex that is within our capabilities.
Noting expressions (9) and (10) and ensuring correct scaling, we take
$$\Gamma_A=\left(\frac{p^2(p+k)^2}{\mu^2k^2}\right)^\gamma\quad {\rm and}\quad
  \Delta_E\Gamma_B=\gamma[p^2-(p+k)^2]\left(\frac{p^2k^2}{\mu^2(p+k)^2}
  \right)^\gamma.$$
Inserting this into (24) we end up with
$$1 = -\frac{a}{2\gamma}\left[1 + \frac{5\gamma}{2} + \frac{13\gamma^2}{4}
 +\cdots\right],\qquad a\equiv \frac{g^2}{16\pi^2}, $$
corresponding to 
\begin{equation}
 \gamma_{\rm model 3}=-\frac{a}{2} + \frac{a^2}{2^3} - \frac{14a^2}{2^5}
 +\cdots
\end{equation}
This result should be compared with the rainbow and chain approximations 
\cite{BK} which, for Yukawa theory, read
$$\gamma_{\rm rainbow}=-\frac{a}{2}+\frac{a^2}{2^3}-\frac{2a^2}{2^5}+\cdots,
\quad\gamma_{\rm chain}=-\frac{a}{2}+\frac{a^2}{2^3}-\frac{2a^2}{2^5}+\cdots$$

\subsection{Lightlike Model 4 for $(\phi^\dag\phi\sigma)_{6D}$}
We saw earlier that in the lightlike limit of one of the momenta, {\em not
necessarily the limit of zero momentum transfer}, the vertex function assumed
the form of the difference of two self-energies, at least to first order in
perturbation theory. Let us therefore make the approximation
$$\Delta_\phi(p+k)\Gamma(p+k,p)\Delta_\phi(p)\simeq\frac{\Delta_\phi(p+k)-
  \Delta_\phi(p)}{p^2-(p+k)^2},$$
in quenched $(\phi^\dag\phi\sigma)_{6D}$ theory, since we have already 
disregarded the $\sigma$ field dressing. Using the usual spectral form of the
$\phi$ field propagator, we may therefore substitute
\begin{equation}
\hspace{-1in}
 \Delta_\phi(p)=\int_0^\infty\frac{\rho(w^2)\,dw^2}{p^2-w^2},\quad
 \Delta_\phi(p+k)\Gamma(p+k,p)\Delta_\phi(p)\simeq
 \int_0^\infty\frac{\rho(w^2)\,dw^2}{((p+k)^2-w^2)(p^2-w^2)}
\end{equation}
in the Schwinger-Dyson equation,
\begin{equation}
 Z_\phi^{-1}= p^2\Delta_\phi(p)-ig^2Z_\phi^{-1}Z_g\int\bar{d}^6k\,
 \Delta_\phi(p+k)\Gamma(p+k,p)\Delta_\phi(p)/k^2.
\end{equation}
Recalling that $Z_\phi^{-1}=\int\rho(w^2)dw^2$, the spectral equation reduces
to 
\begin{equation}
\int dw^2\,\rho(w^2)\frac{w^2Z_\phi Z_g^{-1}+\Sigma(p,w)}{p^2-w^2} = 0,
\end{equation}
where $\Sigma(p,w) = g^2\int\bar{d}^{2\ell}k/[k^2((p+k)^2-w^2)]$ is the first
order self-energy for a $\phi$ field of mass $w$, to be taken in the limit
as $\ell\rightarrow 3$. This has essentially the same form as in the gauge 
technique \cite{DW} for QED, except that it is no longer true that $Z_g=Z_\phi$ 
--- and this is just what one needs! The point is that the self-energy carries
the infinity (in its real part) leaving us with the representation
$$\Sigma(p,w)=\frac{a}{\ell-3}\left[\frac{1}{6}p^2-\frac{1}{2}w^2\right]+\frac
{(p^2-w^2)^2}{\pi}\int\frac{\Im\Sigma(s,w)\,ds}{(s-w^2)^2(s-p^2-i\epsilon)},$$
while $Z_\phi^{-1} Z_g = 1+2a/(3(\ell-3)$ and they both combine neatly
to produce a factor $(p^2-w^2)$ in the numerator of (28).
Therefore taking the imaginary part of (28), one arrives at
\begin{equation}
 -\pi p^2\rho(p^2) + \int\frac{\Im\Sigma(p,w)\rho(w^2)\,dw^2}{p^2-w^2}=0.
\end{equation}
$${\rm Since~}\Im\Sigma(p,w)= g^2(p^2-w^2)^2/6(4\pi)^3p^4,\quad {\rm or}\quad
 \Im\Sigma(p,w)/(p^2-w^2)=a(1-w^2/p^2)^2/6,$$
(29) may be solved by use the ansatz $\rho(w^2)\propto (w^2)^{-1-\gamma}$,
yielding the sought-after relation
$$1 = -\frac{a}{3\gamma(1-\gamma)(2-\gamma)}\quad {\rm or}\quad
\gamma_{\rm model 4}=-\frac{a}{6}+\frac{9a^2}{6^3}-\frac{144a^3}{6^5}\cdots $$
only exact to order $a$.
Evidently $\Delta_\phi(p)$ has the same anomalous dimension as its spectral
function (or imaginary part) $\rho$.
 
All the above model conclusions should be treated with great caution and some
scepticism. We supplied models of the true vertex whose scaling coincided to 
second order with known results for the anomalous dimension, but which possessed
incorrect singularities. It would have been quite easy to change radically 
the results of our calculations by inputting equally plausible vertex ans\"atze.
So all we can purport to have demonstrated to the critical reader is that it is 
possible, {\em in principle}, to regulate the skeleton expansion by using 
nonperturbative propagators and vertices all the while staying in integer 
dimensions. In the end we dare only claim that the scheme outlined in the 
introduction is a viable method for discovering the anomalous scaling properties
of the field theory in question. After all, the skeleton expansion has recently 
proved its worth in a similar context \cite{BDK} and we have no reason to 
suspect that it will fail us in the present circumstances. This said we are
at a computational impasse in our approach in attempting to include the proper 
vertex with its full complement of singularities; unfortunately we see 
no easy way out of this difficulty if we shy away from numerical methods.

\ack
We thank the Australian Research Council for providing financial support 
for this project, under grant number A00000780. We also benefitted from
conversations with Dirk Kreimer and M Kalmykov.

\section*{Appendix}
Here we present a number of auxiliary integrals which assist in determining
the connection between $\gamma$ and $g^2$. The first one of these is
\begin{equation}
 I_{abc}\equiv -i\int\bar{d}^{2\ell}k/(k^2)^a ((p+k)^2)^b \Delta_E^c;
\qquad \Delta_E=4[k^2p^2-(k\cdot p)^2].
\end{equation}
The case $c=0$ is rather well-known to practioners in this field and can 
be found, using Feynman parametric techniques, to equal
\begin{equation}
 I_{ab0}= \frac{(p^2)^{\ell-a-b}}{(-4\pi)^\ell}\frac {\Gamma(a+b-\ell)
 \Gamma(\ell-a)\Gamma(\ell-b)}{\Gamma(a)\Gamma(b)\Gamma(2\ell-a-b)}.
\end{equation}
For $c\neq 0$ we must resort to another method in order to find $I_{abc}$.
Namely we go to the frame where $p=i(\sqrt{q^2};\vec{0}); q^2\equiv -p^2$,
$k=\sqrt{K^2}(i\cos\theta;\sin\theta,\ldots); K^2\equiv -k^2$, so
$$\bar{d}^{2\ell}k = i\frac{K^{2\ell-1}dK}{(2\pi)^{2\ell}}.
 (\sin\theta)^{2\ell-2}d\theta.\frac{2\pi^{\ell-1/2}}{\Gamma(\ell-1/2)},$$
$$(p+k)^2=-q^2-K^2-2\sqrt{q^2K^2}\cos\theta; \qquad
  \Delta_E=4K^2q^2\sin^2\theta.$$
Hence
\begin{equation}
\hspace{-1in}
I_{abc}=\int_0^\infty\frac{2\pi^{\ell-1/2}K^{2\ell-1}\,dK}
    {(2\pi)^{2\ell}\Gamma(\ell-1/2)}\int_0^\pi
 \frac{(\sin\theta)^{2\ell-2}\,d\theta}{(-K^2)^a(-q^2-K^2-2qK\cos\theta)^b
       (4K^2q^2\sin^2\theta)^c}.
\end{equation}
But the standard texts inform us that
$$\int_0^\pi\frac{(\sin\theta)^{\beta-1}\,d\theta}{(1+2z\cos\theta+z^2)^\alpha}
=\sqrt{\pi}\Gamma(\beta)F(\alpha,\alpha-\beta+1/2;\beta+1/2;z^2)
/\Gamma(\beta+1/2)$$
and
$$\int_0^\infty K^{2\sigma-1}F(\alpha,\beta;\gamma;\frac{q^2}{K^2})\,dK
= (-q^2)^\sigma\frac{\Gamma(\gamma)\Gamma(-\sigma)\Gamma(\alpha+\sigma)
   \Gamma(\beta+\sigma)}{2\Gamma(\alpha)\Gamma(\beta)\Gamma(\gamma+\sigma)}.$$
Putting this all together we end up with
\begin{equation}
\hspace{-1in}
 I_{abc}=\frac{(p^2)^{\ell-a-b-2c}}{(-4\pi)^\ell 4^c}\frac{\Gamma(\ell-c-1/2)}
   {\Gamma(\ell-1/2)}\frac{\Gamma(a+b+c-\ell)\Gamma(\ell-a-b)\Gamma(\ell-b-c)}
   {\Gamma(a)\Gamma(b)\Gamma(2\ell-a-b-2c)}.
\end{equation}
In particular, for 6D, the result reads
\begin{equation}
 I_{ab1}=\frac{(p^2)^{1-a-b}}{6(-4\pi)^3}\frac{\Gamma(a+b-2)\Gamma(2-a)
  \Gamma(2-b)}{\Gamma(a)\Gamma(b)\Gamma(4-a-b)}.
\end{equation}
Therefore if one takes $b$ non-integer for the present, the limit as
$a\rightarrow 0, -1$ will produce a vanishing result leading to
$I_{0b1} = I_{-1b1}=0$. We take this result to be correct even for 
integer $b$, paralleling the treatment of tadpole integrals in dimensional
regularization.

\end{document}